\documentclass[conference]{IEEEtran}
\usepackage{graphicx}
\usepackage{bm}
\usepackage{cite}
\usepackage{amsmath}
\usepackage{footnote}

\newcommand\aj{{\it Astron. J.}}

\newcommand\apj{{\it Astrophys. J.}}
\newcommand\apjl{{\it Astrophys. J. Lett.}}
\newcommand\apjs{{\it Astrophys. J.Supp.}}

\newcommand\mnras{{\it Mon. Not. R. Astron. Soc.}}
\newcommand\nat{{\it Nature}}
\newcommand\prd{{\it Phys, Rev, D.}}
\newcommand\pasj{{\it Publ. of the Astron. Society of Japan}}
\newcommand\na{{\it New Astronomy}}

\begin{document}

\title
{
4.45 Pflops Astrophysical $N$-Body Simulation on K computer - The Gravitational Trillion-Body Problem
}

\author{\IEEEauthorblockN{Tomoaki Ishiyama}
\IEEEauthorblockA{ 
Center for Computational Science \\
University of Tsukuba \\
ishiyama@ccs.tsukuba.ac.jp}
\and
\IEEEauthorblockN{Keigo Nitadori}
\IEEEauthorblockA{
Center for Computational Science \\
University of Tsukuba \\
keigo@ccs.tsukuba.ac.jp} 
\and
\IEEEauthorblockN{Junichiro Makino}
\IEEEauthorblockA{
Graduate School of Science and Engineering \\
Tokyo Institute of Technology\\
makino@geo.titech.ac.jp
}}

\date{}

\maketitle


\begin{abstract}
As an entry for the 2012 Gordon-Bell performance prize, we report
performance results of astrophysical $N$-body simulations of one
trillion particles performed on the full system of K computer.  This
is the first gravitational trillion-body simulation in the world.  
We describe the scientific motivation, the
numerical algorithm, the parallelization strategy, and the performance
analysis.  Unlike many previous Gordon-Bell prize winners that used
the tree algorithm for astrophysical $N$-body simulations, we used the
hybrid TreePM method, for similar level of accuracy in which the short-range
force is calculated by the tree algorithm, and the long-range force is
solved by the particle-mesh algorithm.  We developed a highly-tuned
gravity kernel for short-range forces, and a novel communication
algorithm for long-range forces.  The average performance on 24576 and
82944 nodes of K computer are 1.53 and 4.45 Pflops, which
correspond to 49\% and 42\% of the peak speed.
\end{abstract}

\section{Introduction}

Astrophysical $N$-body simulations have been widely used to study the
nonlinear structure formation in the Universe. Such simulations are
usually called as cosmological $N$-body simulations.  In these
simulations, a particle moves according to the gravitational forces
from all the other particles in the system.

Thus, the most straightforward algorithm to calculate the acceleration
of a particle is to calculate the $N-1$ forces from the rest of the
system, where $N$ is the total number of particles in the system. This
method is usually called the direct summation. This method is
unpractical for large $N$ ($N>10^6$), since the calculation cost is
proportional to $N^2$. Therefore, faster algorithms with some
approximation are usually used in cosmological $N$-body simulations.

The tree algorithm \cite{Barnes1986}\cite{Barnes1990} is the most
widely used algorithm for cosmological $N$-body simulations.  The
basic idea of the tree algorithm is to use a hierarchical oct-tree
structure to represent an $N$-body system (figure \ref{fig:tree}). The
force from particles in one box to one particle can be calculated by
evaluating the multipole expansion, if the error is small enough (if
the box and the particles are well separated). If not, the force is
evaluated as the sum of forces from eight subboxes. By recursively
applying this procedure, one can calculate the total force on a
particle with $\mathcal O(\log N)$ cost, 
{\footnotesize \\\\SC12, November 10-16, 2012, Salt Lake City, Utah, USA \\
978-1-4673-0806-9/12\$31.00 \copyright 2012 IEEE\\}\noindent 
and the total calculation cost per
timestep becomes $\mathcal O(N\log N)$, achieving 
drastic reduction from the
$\mathcal O(N^2)$ cost of the direct summation method.

Thus, the tree algorithm have been used for most of practical large cosmological
calculations in the last two decades, and many Gordon-Bell prizes have
been given to such simulations (1992\cite{Warren1992},
1995-2001
{\cite{Makino1995,Fukushige1996,Warren1997,Warren1998,Kawai1999,Makino2000,
  Makino2001}, 2003\cite{Makino2003}, and
2009-2010\cite{Hamada2009}\cite{Hamada2010}).  In order to accelerate
the calculation of gravity, some of them used GRAPEs
\cite{Makino1995}\cite{Fukushige1996}\cite{Kawai1999,Makino2000,Makino2001,Makino2003}
, which are special-purpose computers
\cite{Sugimoto1990,Makino1998,Makino2003b}, or graphics processing
units (GPUs) \cite{Hamada2009}\cite{Hamada2010}.

There is one difference between large cosmological simulations in the
literature and those awarded Gordon-Bell prizes so far. In practically
all recent large calculations, except for those for Gordon-Bell
prizes, the periodic boundary condition is used. With the periodic
boundary condition, the computational domain is a cube, and we assume
that there are infinite copies of them which fill the infinite
space. This condition is used to model the Universe which is uniform
in very large scale.  On the other hand, in simulations for past
Gordon-Bell prizes, the open boundary condition, in which the
calculation domain is initially a sphere, was used.

\begin{figure}[!t]
\centering
\includegraphics[width=8cm]{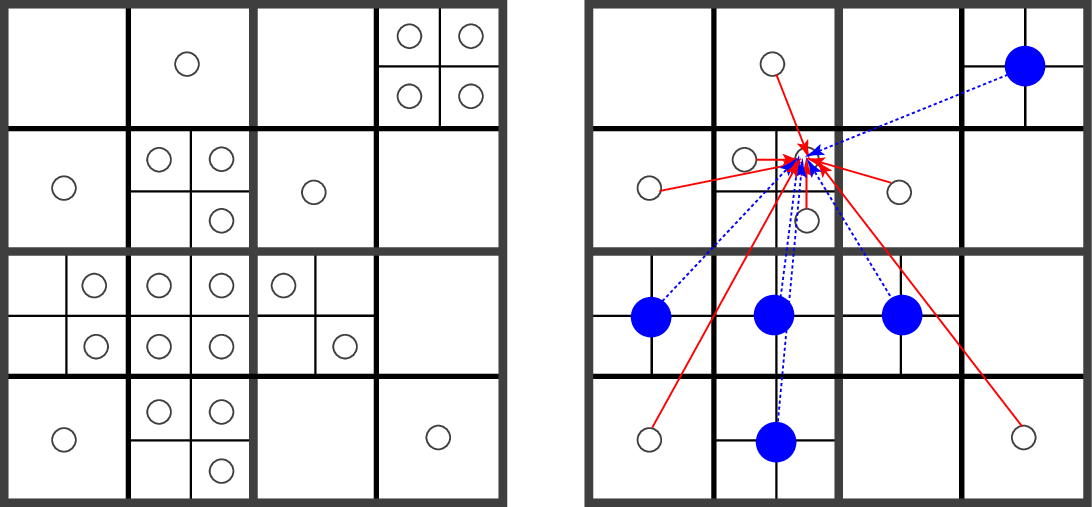}
\caption{ The hierarchical tree algorithm.  White circles represent
  particles.  Blue circles are the multipole expansions of tree nodes.
  Red solid arrows and blue dotted arrows show the particle-particle
  and the particle-multipole interactions, respectively.  }
\label{fig:tree}
\end{figure}

\begin{figure}[!t]
\centering
\includegraphics[width=8cm]{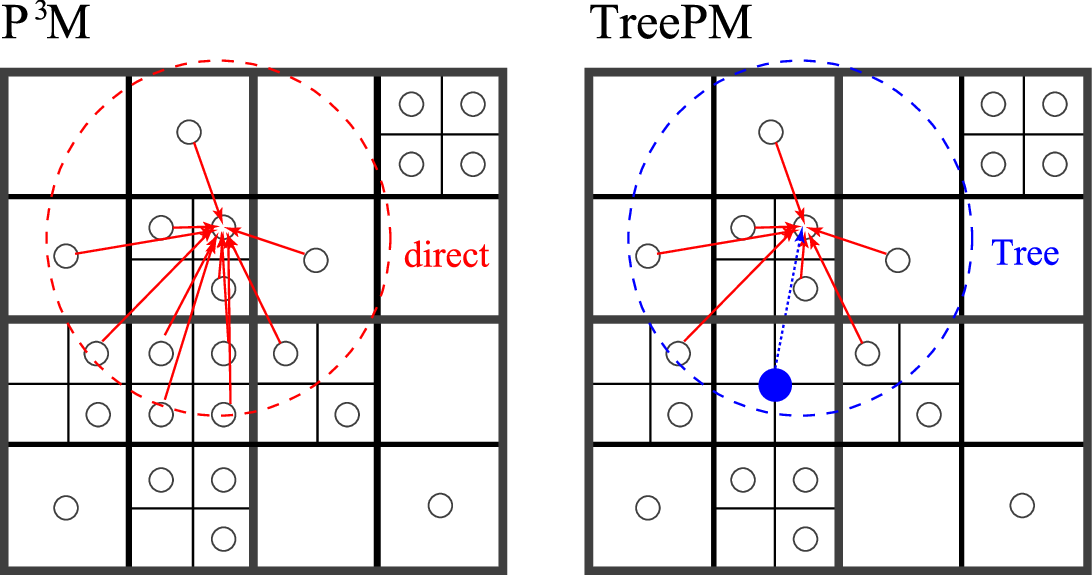}
\caption{ The schematic view of the $\rm P^3M$ and the TreePM
  algorithm.  Large red and blue dashed circles show the cutoff radius
  of each algorithm.  Within these radii, the short-range forces are
  calculated by the direct method or the tree algorithm.  White
  circles represent particles.  Blue circles are the multipole
  expansions of tree nodes.  Red solid arrows and blue dotted arrows
  show the particle-particle and the particle-multipole interactions,
  respectively.  The residual force is calculated
  by the PM algorithm.  }
\label{fig:treepm}
\end{figure}

From theoretical point of view, in both cases one can only analyze
structures of the size sufficiently small compared to the size of the
computational domain. However, periodic boundary is computationally
more efficient. The reason is that with open boundary, only the
structures near the center of the sphere are reliable. Structures near
the boundary are affected by the presence of the boundary to the
vacuum.  Thus, only a small fraction of the total computational volume
is useful for the analysis of the structure formed. On the other hand,
with the periodic boundary, everywhere is equally reliable and can be
used for the analysis.

The PM (Particle Mesh) algorithm has been widely used in 
cosmological $N$-body simulations with the periodic boundary
condition since 1980's. The PM algorithm can obtain the gravitational
potential on a regular grid.  The mass density at a grid point
is calculated by assigning the masses of nearby particles by
some kernel function.  
Then, the Poisson equation is solved using FFT.
Finally, the
gravitational force on a particle position is obtained by
differentiating and interpolating the potential on the mesh.  
For details, see Eastwood (1981) \cite{Hockney1981}.

In general, the PM algorithm is much faster but less accurate than the
tree algorithm since the spatial force resolution is limited by the
size of the mesh.  In order to overcome this problem, hybrid algorithm
such as the $\rm P^3M$ (Particle-Particle Particle-Mesh) and the
TreePM (Tree Particle-Mesh) algorithm have been developed.  The main
idea of these algorithm is that the gravitational force is split into
two components, short- and long-range forces.  The short-range force
decreases rapidly at large distance, and drops zero at a finite
distance.  This part with the cutoff function on the force shape is
calculated by a high resolution algorithm, such as the direct summation
($\rm P^3M$) or the tree algorithm (TreePM) (e.g. \cite{Xu1995,
  Bode2000, Bagla2002, Dubinski2004, Springel2005b, Yoshikawa2005,
  Ishiyama2009b}).  The long-range force drops at large wavenumbers in
the frequency space, and is calculated by the PM algorithm.  Figure
\ref{fig:treepm} shows a schematic view of the $\rm P^3M$ and the
TreePM algorithm.

In the $\rm P^3M$ and TreePM algorithms, we can calculate the
gravitational potential with high spatial resolution under the
periodic boundary condition.  In cosmological $N$-body simulations,
structures with high density form rapidly from the small initial
density fluctuations via the gravitational instability.  It is not
practical to use the $\rm P^3M$ algorithm since the computational cost
of the short-range part increases rapidly as the formation proceeds.
The calculation cost of a cell within the cutoff radius with $n$
particles is $\mathcal O(n^2)$.  Thus, for a cell with 1000 times more
particles than average, the cost is $10^6$ times more expensive.  The
TreePM algorithm can solve this problem, since the calculation cost of
such cell is $\mathcal O(n\log n)$.  Thus, the TreePM algorithm is
very efficient and is used in a number of recent large cosmological
$N$-body simulations (e.g. \cite{Springel2005} \cite{Ishiyama2009}).

The TreePM algorithm is relatively new. It became popular around 2000.
Thus, it is not surprising that Gordon-Bell winners in 1990's used
the pure tree algorithm.  Advantage of the TreePM algorithm over the pure
tree algorithm is twofold. The first one is, as we already noted, we
can use entire volume for data analysis.  The second one is that for
the same level of accuracy, the TreePM algorithm requires
significantly less operations.  With the tree algorithm, the
contributions of distant (large) cells dominate the error in the
calculated force.  With the TreePM algorithm, the contributions of
distant particles are calculated using FFT. Thus, we can allow relatively
moderate accuracy parameter for the tree part,
resulting in considerable reduction in the computational cost.

In this paper, we describe our MPI/OpenMP hybrid TreePM implementation
GreeM \cite{Ishiyama2009b}, which is a massively parallel TreePM code
based on the implementation of Yoshikawa \& Fukushige (2005)
\cite{Yoshikawa2005} for large cosmological $N$-body simulations, and
present performance results of the largest gravitational $N$-body
simulation that has ever been done in the world.  We use one trillion
dark matter particles to study the nonlinear evolution of the first
dark matter structures in the early Universe.  The numerical
simulations were carried out on K computer at the RIKEN Advanced
Institute for Computational Science, which is the world's fastest
supercomputer at the time this paper is submitted.  It consists of
82944 SPARC64 VIIIfx oct-core processors with the clock speed of 2.0
GHz (the total number of core is 663552) and 1.3PB of memory. The peak
performance is 10.6 Pflops.

The scientific motivation of this simulation is to study the nature
of dark matter particles, which is a long-standing problem in both
astrophysics and  particle physics.  One candidate of 
the dark matter particle is the lightest supersymmetric particle, the
neutralino.  Since the neutralino is itself its anti-particle, it
self-annihilates and produces new particles and gamma-rays. Indirect
detection experiments to detect these productions are the important
way to study the nature of dark matter.  Therefore, such indirect
detection experiments are one of the key projects of the current
generation gamma-ray space telescope Fermi \cite{Abdo2009} and the next
generation ground-based chelenkov telescope array
CTA\footnote{http://www.cta-observatory.org/}.

The Milky Way is in the bottom of the potential of a dark matter halo, 
which is ten times larger than the Milky Way itself.  For
precise predictions of the annihilation gamma-rays in the Milky Way,
we need to know the fine structures of dark matter halos, since the
gamma-ray flux from dark matter structures is proportional to the
square of their density and inverse square of the distance from us.
Ishiyama et al. (2010) \cite{Ishiyama2010} found that the central
density of the smallest dark matter structures is very high.  If they
survive near our Sun, the annihilation signals could be observable as
gamma-ray point-sources.  Thus, the behavior of the smallest dark
matter structures near Sun is very important for the indirect
detection experiments.

The cosmological $N$-body simulation is useful for the study of the
structure and evolution of dark matter structures within the Milky
Way.  Analytical studies (e.g. \cite{Zybin1999}\cite{Berezinsky2003})
predicted the mass of the smallest structures to be comparable to the
mass of the earth, which is 18 orders of magnitude smaller than
that of the Milky Way.  Unfortunately, we cannot simulate such a wide
dynamic range with currently available computational resources.
Ishiyama et al. 2010\cite{Ishiyama2010} simulated only the smallest
structures.  With the simulation described here, we extend this
strategy further.  We simulate a much larger volume than that used in
Ishiyama et al. 2010\cite{Ishiyama2010}, and study the evolution.
Using these results, we can predict their evolution of the smallest
structures survive or not near Sun.

This paper is organized as follows.  First, we describe our parallel
TreePM implementation.  Then we present the detail of the simulation
with one trillion particles and report the performance of our code on
K computer.

\section{Our Parallel TreePM Code}
\newcommand{\rcut}{r_{\rm cut}}

In the TreePM method, the force on a particle is divided into two
components, the long-range (PM) part and the short-range (PP:
particle-particle) part. The long-range part is evaluated by FFT, and
the short-range part is calculated by the tree method, with a 
cutoff function on the force shape.
The density of point mass $m$ is decomposed into 
PM part and PP part:
\begin{eqnarray}
\label{eq:density}
\rho_{\rm PM}(r) &=& \left\{
\begin{array}{l}
{\displaystyle \frac{3m}\pi 
\left(\frac{2}{\rcut}\right)^3 \left( 1 - \frac{r}{\rcut/2} \right),} \vspace{3mm} \\
{\displaystyle \qquad \qquad \quad (0 \le r \le \rcut/2), } \vspace{3mm}\\ 
{\displaystyle 0, \qquad \qquad \qquad (r >\rcut/2), } \nonumber
\end{array}
\right.  \\ \\ \nonumber
\rho_{\rm PP}(r) &=& \frac{m}{4\pi} \delta^3(r) - \rho_{\rm PM}(r).
\end{eqnarray}
The function $\rho_{\rm PM}$ expresses a linearly 
decreasing density (shape S2) \cite{Hockney1981}.  Since $4\pi\int_0^{\rcut/2}
r^2 \rho_{\rm PM}(r)dr = m$ and $4\pi\int_0^{\rcut/2} r^2 \rho_{\rm
  PP}(r)dr = 0$, the particle-particle interaction vanishes outside the finite
radius $\rcut$ (Newton's second theorem). We use a small
softening with length $\varepsilon \ll \rcut$ to the short-range
interaction that corresponds to replacing the delta function with a
small kernel function.

Given the positions $\bm r_i$, the masses $m_i$ and the gravitational
constant $G$, the particle-particle interaction takes the form
\begin{eqnarray}
\bm f_i = \sum_{j \neq i}^N G m_j \frac{\bm r_j - \bm r_i}{|\bm r_j -
  \bm r_i|^3} g_{\rm P3M}(2|\bm r_j - \bm r_i|/\rcut).
\end{eqnarray}
The cutoff function $g_{\rm P3M}$ 
is obtained by evaluating the force between two particles with the density of 
equation (\ref{eq:density}) by six-dimensional spatial integration,
\begin{eqnarray}
\label{eq:cutoff}
g_{\rm P3M}(\xi)=\left\{
\begin{array}{l}
{\displaystyle 1 + \xi^3(-\tfrac{8}{5} + \xi^2(\tfrac{8}{5}
+ \xi(-\tfrac{1}{2} + \xi(-\tfrac{12}{35} + \xi\tfrac{3}{20})))) } \vspace{3mm}\\
{\displaystyle - \zeta^6 (\tfrac{3}{35} + \xi(\tfrac{18}{35} + \xi\tfrac{1}{5})), 
\quad (0 \le \xi \le 2),} \vspace{3mm}  \\
{\displaystyle \qquad \qquad \quad {\rm where} \quad \zeta = \max(0, \xi-1) },  \vspace{3mm}\\
{\displaystyle 0, \quad (\xi>2)}.  \\
\end{array}
\right.
\end{eqnarray}
Here, we modified the original form \cite{Hockney1981} with a branch at $\xi=1$
, which is optimized for the evaluation on a 
SIMD (Single Instruction Multiple Data) hardware with 
FMA (Fused Multiply-Add) support.

For the tree part, we used Barnes' modified algorithm
\cite{Barnes1990} in which the traversal of the tree structure is done
for a group of particles.  It is done for each particle in the original
algorithm \cite{Barnes1986}.  
In the modified algorithm, 
a list of tree nodes and particles are
shared by a group of particles.  The forces from nodes and particles
in the list to particles in groups are calculated directly.  This
modified algorithm can reduce the computational cost of tree traversal
by a factor of $\langle N_i \rangle$, where $\langle N_i \rangle$ is
the average number of particles in groups.  On the other hand, the
computational cost for the PP force calculation increases since the
interactions between particles in groups are calculated directly, and
the average length of the interaction list becomes longer. The optimal
value of $\langle N_i \rangle$ depends on the performance
characteristics of the computer used.  It is around 100 for K
computer, and 500 for a GPU cluster \cite{Hamada2009}.

We use a 3-D multi-section decomposition \cite{Makino2004}.  As a
result, the shape of a domain is rectangular.  We use the sampling
method \cite{Blackston1997} to determine the geometries of domains.
The sampling method allows us to drastically reduce the amount of
communication needed for constructing the division because we use only
a small subset of particles. It is difficult to achieve good load
balance for the following reason.  In cosmological $N$-body
simulations, the initial particle distribution is close to uniform
with small density fluctuations.  These fluctuations grow nonlinearly
via the gravitational instability and form a number of high density
dark matter structures in the simulation box.  The density of such
structures are typically a hundred or a thousand times higher than the
average.  Sometimes, the central density of these structures can reach
$\sim$$10^7$ times the average.  In such a situation, the calculation
cost of the short-range part becomes highly imbalanced,
if the domain decomposition is static, in other words, its geometry of
each domain is time invariable and is the same for all domains.

In our method, we adjust the geometries of the domains assigned to
individual processes, so that the total calculation time of the force
(sum of the short-range and long-range forces) becomes the same for
all MPI processes.  We achieve this good load balance by adjusting the
sampling rate of particles in one domain according to their
calculation costs.  We adjust the sampling rate of particles in one
domain so that it is proportional to the measured calculation time of
the short-range and long-range forces.  Thus, if the calculation time
of a process is larger than the average value, the number of sampled
particles of the process becomes relatively larger.  After the root
process gathers all sampled particles from the others, the new domain
decomposition is created so that all domains have the same number of
sampled particles.  Therefore, the size of the domain for this process
becomes somewhat smaller, and the calculation time for the next
timestep is expected to become smaller.

Figure \ref{fig:domain} shows the domain decomposition for a
cosmological simulation.  We can see that high density structures are
divided into small domains so that the calculation costs of all
processes are the same.

We update the geometries every step following the evolution of the
simulated system.  The cost is negligible compared to the
effect of the load imbalance.  However, often large
jumps of boundaries occur since there are fluctuations due to
sampling.  In order to avoid the large jumps of the boundaries, we
adopt the linear weighted moving average technique for boundaries of
last five steps.  Thus, we suppress sudden increment of the amount of
transfer of particles across boundaries.

The detailed explanation of the GreeM code can be found in Ishiyama et
al. (2009) \cite{Ishiyama2009b}.  In the rest of this section, we
describe two novel techniques that significantly improved the
performance. The first is the near-ultimate optimization of the
equation (\ref{eq:cutoff}).  The second is the relay mesh method for
the PM part.

\subsection{Optimized Particle-Particle Force Loop}\label{sec:kernel}
Most of the CPU time is spent for the evaluation of the
particle-particle interactions. Therefore we have developed a highly
optimized loop for that part. This force loop was originally developed
for the x86 architecture with the SSE instruction set, and named
Phantom-GRAPE\cite{Nitadori2006,Tanikawa2012,Tanikawa2012b} after its
API compatibility to GRAPE-5 \cite{Kawai2000}.  We have ported
Phantom-GRAPE with support for the cut-off function
(eq.\ref{eq:cutoff}) to the HPC-ACE architecture of K computer.

The LINPACK peak per core of SPARC64 XIIIfx is 16 Gflops
(4 FMA units running at 2.0 GHz). However, the theoretical upper limit
of our force loop is 12 Gflops because it consists of 17 FMA and
17 non-FMA operations ($51 \times 2$ floating-point operations in total)
\footnote{ 17 {\tt fmadd/fmsub/fnmadd/fnmsub}, 9 {\tt fmul}, 4 {\tt
    fadd/fsub}, 3 {\tt fmax/fcmp/fand}, and 1 {\tt frsqrta}.  } for
two (one SIMD) interactions.  Our force loop reaches 11.65 Gflops on a
simple $\mathcal O(N^2)$ kernel benchmark, which is 97\% of the
theoretical limit.

We have implemented the force loop with SIMD built-in functions
provided by the Fujitsu C++ compiler which define various operations
on a packed data-type of two double-precision numbers.  The loop was
unrolled eight times by hand so that 16 pairwise interactions, forces from
4-particles to 4-particles are evaluated in one iteration.
Furthermore the loop was ten-times unrolled and software-pipelined by
the compiler.

An inverse-square-root was calculated using a fast approximate
instruction of HPC-ACE with 8-bit accuracy and a third-order
convergence method with $y_0 \sim 1/\sqrt{x},\ h_0 = 1 - xy_0^2,\ y_1
= y_0(1 + \frac12 h_0 + \frac38 h_0^2)$ to obtain 24-bit accuracy.  A
full convergence to double-precision will increase both CPU time and
the flops count, without improving the accuracy of scientific results.

\begin{figure}[!t]
\centering
\includegraphics[width=8cm]{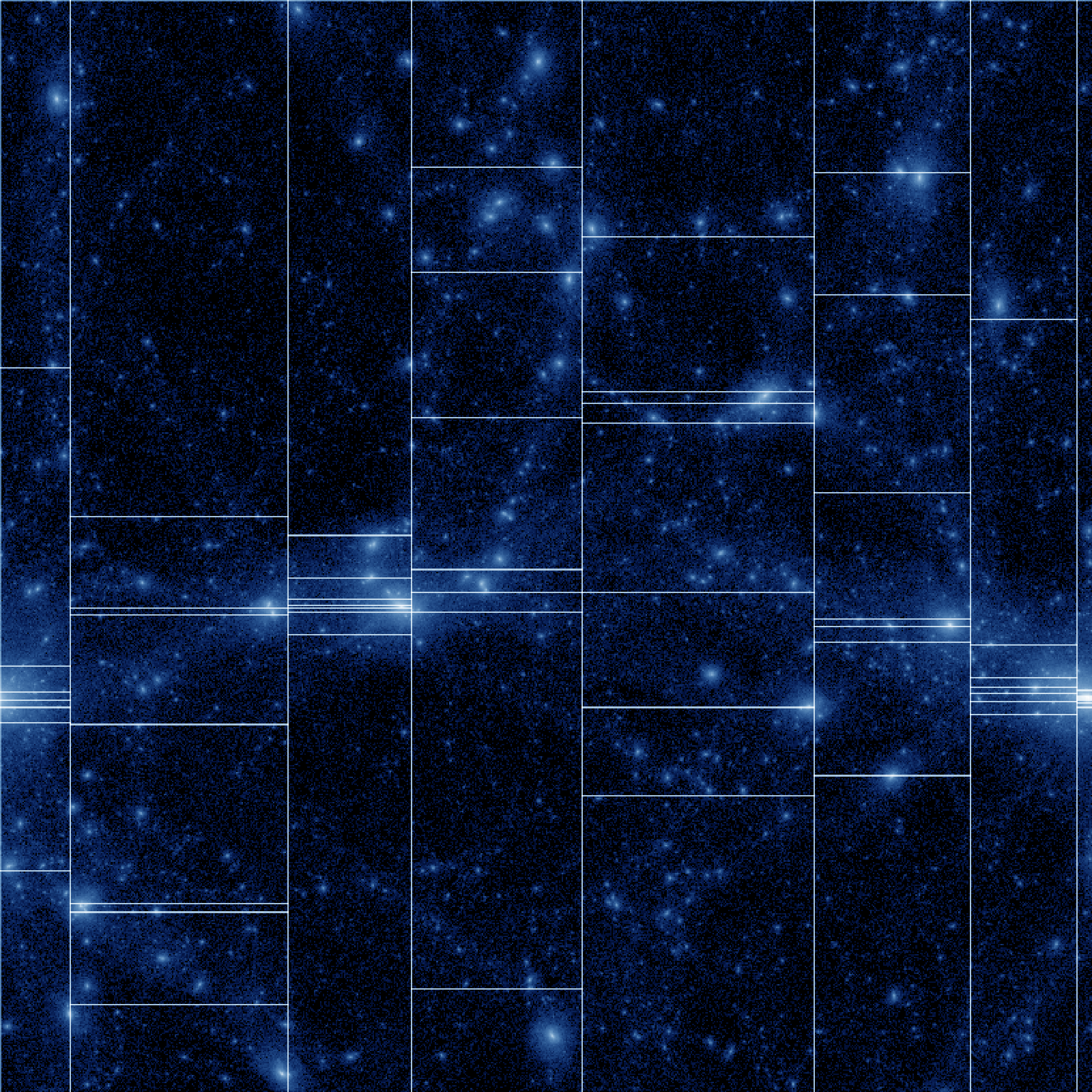}
\caption{
The example of the domain decomposition.
It shows $8 \times 8$ division in two dimensions.
}
\label{fig:domain}
\end{figure}

\subsection{Relay Mesh Method}

\begin{figure}[!t]
\centering
\includegraphics[width=8cm]{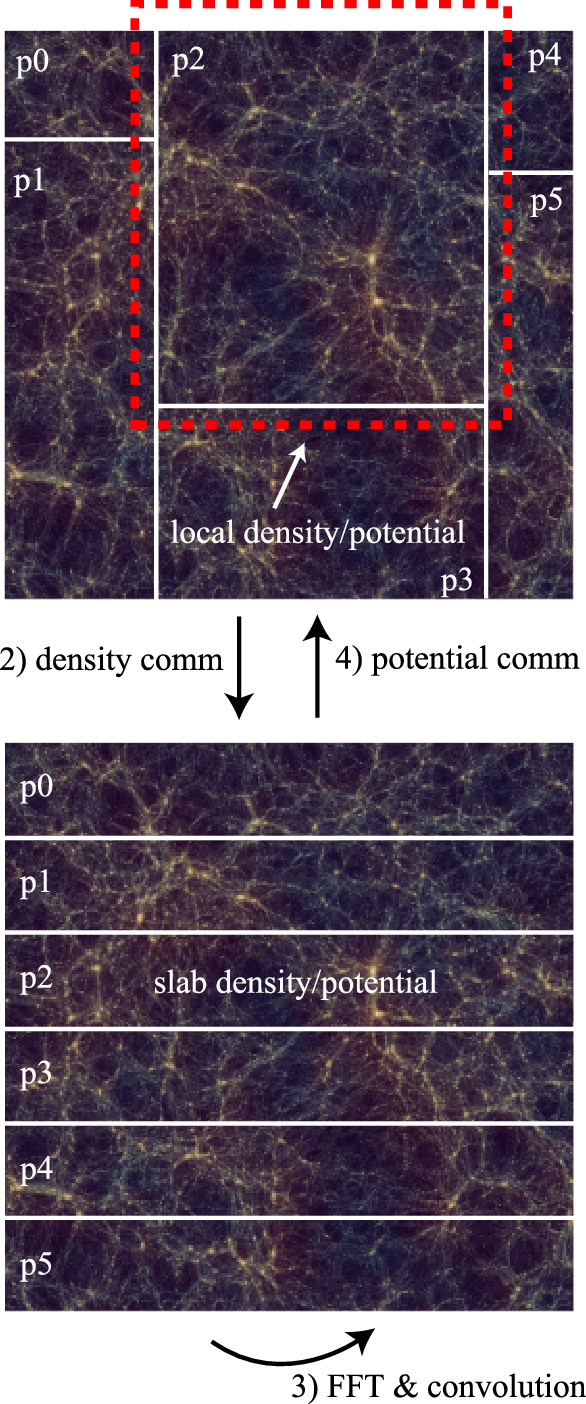}
\caption{ Two domain decompositions for the PM method.  Upper panel
  shows the domain decomposition of local mesh structures.  Bottom
  panel shows the domain decomposition of slab mesh structures for the
  parallel FFT.  In this figure, p[0-5] mean the identification number
  of each process.  The local mesh of a process covers only own domain
  but contains some ghost layer which is needed according to an
  adopted interpolation scheme for the density assignment and the
  calculation of forces on particle positions.  }
\label{fig:pm_parallel}
\end{figure}

For the parallel FFT of the PM part, we use the MPI version of the
FFTW 3.3 library\footnote{http://www.fftw.org/}.  The data layout
supported by the parallel FFTW is the 1-D slab decomposition only.
The drawback of this 1-D parallel FFT is that the number of processes
that perform FFT is limited by the number of grid point of the PM part
in one dimension.  We usually use the number of PM mesh $N_{\rm PM}$
between $N/2^3$ and $N/4^3$ in order to minimize the force error
\cite{Ishiyama2009b}.  If we perform a $10240^3$ particles simulation,
the number of mesh is between $2560^3$ and $5120^3$.  Consequently,
the number of processes perform FFT is $2560\sim5120$, 
which is very small fraction of the total number of cores of K computer.
When we perform
parallel FFT via {\tt MPI\_COMM\_WORLD}, FFTW allocates processes to
perform FFT in ascending order of their ranks in the communicator 
{\tt MPI\_COMM\_WORLD}.  In this case, processes with their ranks
$0\sim2559$ or $0\sim5119$ perform FFT.  The others with larger ranks
do not perform FFT.  Usually, it is not clear how an MPI rank
corresponds to the physical layout of a node.  Thus, when the number
of processes is larger than the number of grid point in one dimension,
if we use {\tt MPI\_COMM\_WORLD}, the communication pattern within the
FFT processes is likely to be not optimized.  In order to avoid this
problem, we select processes to perform FFT so that their physical
positions are close to one another and create a new communicator {\it
  COMM\_FFT} by calling {\tt MPI\_Comm\_split}, which includes these
processes only.

In the following, we first describe a straightforward implementation
and its limitation. Then we describe our solution.  The 1-D parallel
FFT requires the conversion of data layout in order to perform the 1-D
parallel FFT.  Particles are distributed in 3-D domain decomposition
in our parallelization method for the optimization of the load
balance.  This means that optimal domain decompositions are quite
different for the particles and the PM mesh.  However, the conversion
of the data layout of particles would be heavy task since almost all
particles have to be exchanged.  The best way to perform the
conversion is to exchange the mass density on the mesh after the
assignment of the mass of all particles to the local mesh.  After the
assignment, a process has the mesh that covers only its own domain.
Then, each process communicates the local mesh so that the FFT
processes receive the complete slabs.  After the calculation of the
potential on the mesh is completed, we perform the conversion of the
1-D distributed potential slabs to the 3-D distributed rectangular
mesh for the calculation of forces.  Figure \ref{fig:pm_parallel}
shows an illustration of two different domain decompositions for the
PM method.

In GreeM, a cycle of the parallel PM method proceeds in the
following five steps.
\begin{enumerate}
\item 
Each process calculates the mass density on the local mesh by
assigning the mass of all particles using the TSC (Triangular Shaped
Cloud) scheme, where a particle interacts with 27 
grid points.
The mesh of each process covers only its own domain.
\item 
The FFT processes construct the slab mesh by incorporating the
contributions of the particles in other processes.  Each process sends
the density of the local mesh to other processes by calling
{\tt MPI\_Alltoallv(..., MPI\_COMM\_WORLD)}
\footnote{ One may imagine replacing this communication with {\tt
    MPI\_Isend} and {\tt MPI\_Irecv}.  However, a FFT process
  receives meshes from $\sim$4000 processes.  Such a large number of
  non-blocking communications do not work concurrently.  }.  Each
FFT process receives and sums up only the contributions that overlap
with their own slabs from other processes.
\item
Then the gravitational potential on the slabs are calculated by using
the parallel FFT (via {\it COMM\_FFT}) and performing the convolution
with the Green's function of the long-range force.
\item
The FFT processes send the gravitational potential on the slabs to
all other processes by calling {\tt MPI\_Alltoallv(...,MPI\_COMM\_WORLD)}.  
Each process receives the contributions
that cover its own local mesh from the FFT processes.
\item 
Each process has the gravitational potential on the local mesh that
covers the domain of own process.  The gravitational forces on the
local mesh are calculated by the four point finite difference
algorithm from the potential. Then we calculate the PM forces on
particles by interpolating forces on the local mesh.
\end{enumerate}

\begin{figure*}[!t]
\centering
\includegraphics[width=16cm]{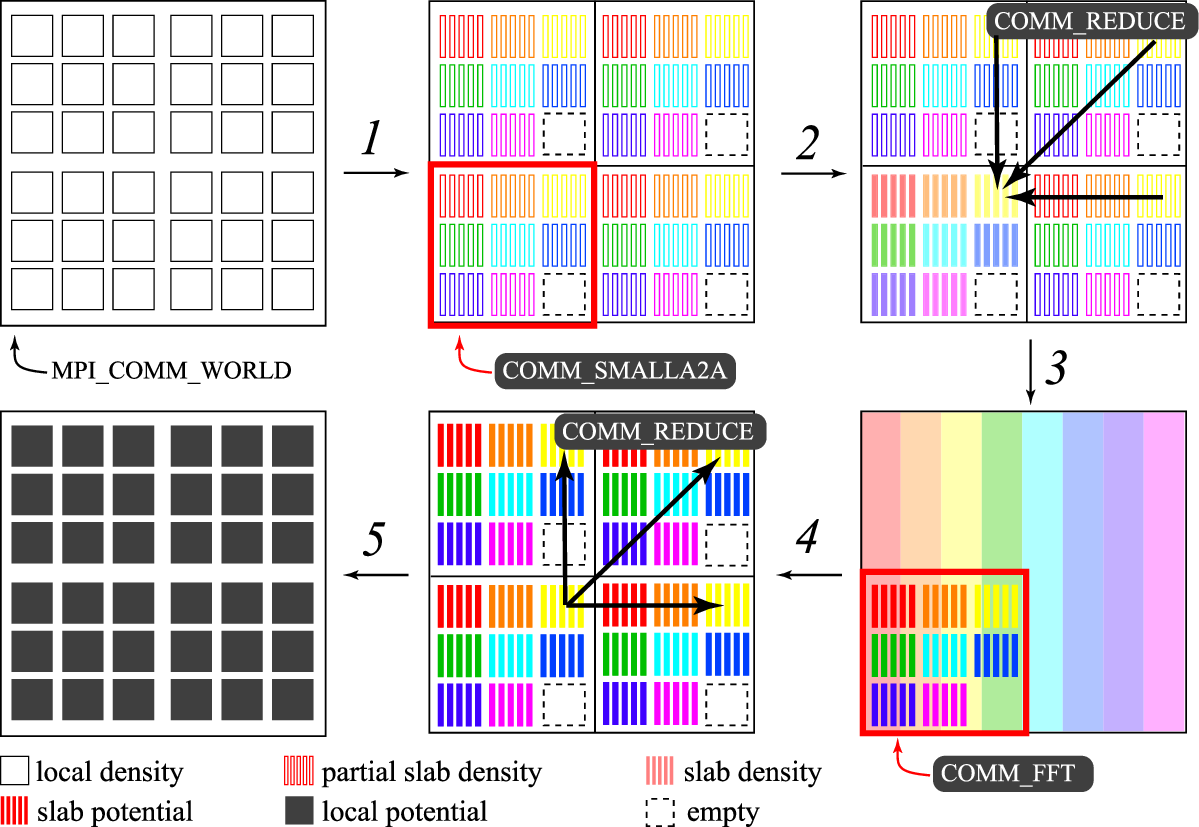}
\caption{
Illustration of our communication algorithm,
relay mesh method. 
There are 2-D decomposed $6 \times 6$ processes. 
The number of the PM mesh is $N_{\rm PM}=8^3$, and 
that of processes perform FFT is eight.
There are four groups that include nine processes.
The detail explanation is written in the text.
The background of bottom-right panel shows 
the physical regions that are corresponding to 
the slab density of each process in the root group. 
}
\label{fig:relaymesh}
\end{figure*}

Since the calculation cost of FFT is relatively small in most cases,
this 1-D parallel FFT does not cause significant degradation of the
performance.  However, in the situation that the number of MPI
processes is very large, communication becomes problematic since the
number of processes that send the local mesh to an FFT process is
proportional to $p^{2/3}$, where $p$ is the number of MPI
processes. When we use 82944 processes, an FFT process receives slabs
from $\sim$4000 processes, and network congestion would occur on the
communication network.  Thus, the communication time of the conversion
of the mesh structures can become bottlenecks on modern massively
parallel environments such as the full system of K computer.

In order to solve this problem, we developed a novel communication
algorithm, {\it Relay Mesh Method}.  The basic idea of this method is
to split the global all-to-all communication on the conversion of the
mesh structures into two local communication.  Processes are divided
into small groups whose sizes are equal or larger than that of the FFT
processes.  One of the groups contains the FFT processes, we call this
group the root group.  For example, consider a simulation with 2-D
decomposed $6 \times 6$ processes and $N_{\rm PM}=8^3$.  In this case,
the number of FFT processes is eight since the FFT is parallelized for
only one axis.  We make four groups that consist of $3 \times 3 = 9$
processes. The eight processes of the root group perform FFT.  The 1-D
slab decomposed density mesh is constructed in the following two
steps.  First, each group compute the contribution of its particles to
the mesh, and then the total mesh is constructed by adding up the
contributions from all groups.
the global communication in the second step (previous page) is
replaced by two local communications, one within groups and the other
over groups.  The first communication is done to construct the 1-D
distributed density mesh in the same way as the second step of the
original method, but the communication is closed within each group.
In this example, the nine processes of each group send the mass
density to eight processes within the same group.  After the first
communication, each group has the 1-D distributed partial density
slabs.  Then, all groups {\it relay} the partial slabs to the root
group, and the root group reduces them to construct the complete
slabs.  In this example, four processes in different groups
communicate.  Figure \ref{fig:relaymesh} shows a schematic view of
this method.

In order to perform these two communication steps, 
we create two communicators {\it COMM\_SMALLA2A} and {\it COMM\_REDUCE}
by calling {\tt MPI\_Comm\_split}.
\begin{itemize}
\item
{\it COMM\_SMALLA2A} : Each process in a group can communicate with
each other through this communicator (showed in the small red box
in figure \ref{fig:relaymesh}).
One of group includes FFT processes (the root group).
\item 
{\it COMM\_REDUCE} : Each process can communicate with processes in other
groups with the same rank in the communicator {\it COMM\_SMALLA2A}
(showed in same colors in figure \ref{fig:relaymesh}).
The number of processes in this communicator is same as the 
number of groups.
\end{itemize}
In the case of the example showed in figure \ref{fig:relaymesh}, 
the number of groups is four, 
and that of processes in each group is nine.
Eight processes of the root group perform FFT.
The number of communicater {\it COMM\_REDUCE} is eight, 
in which there are four processes.

The PM procedures 2-4 explained early in this subsection are
replaced by following five steps 
(the numbers correspond to that in figure \ref{fig:relaymesh}).
\begin{enumerate}
\item
After the density assignment, each process sends the density of the
local mesh to other processes in its group by calling
{\tt MPI\_Alltoallv(..., {\it COMM\_SMALLA2A})}.  In the case of figure
\ref{fig:relaymesh}, the 3-D distributed local mesh of nine
processes are communicated to eight processes as the latter has the
1-D distributed partial density slabs.
\item
Each process sends the slabs to the corresponding process in the root
group by using {\tt MPI\_Reduce(..., {\it COMM\_REDUCE})}.  After this
communication, each process of the root group has the 1-D distributed
complete slabs.
\item
In the FFT processes, the gravitational potential on the slabs are
calculated by using parallel FFT (via {\it COMM\_FFT}).
The processes in other groups wait the end of FFT.
\item
The root group sends the slab potential to other groups by means of
{\tt MPI\_Bcast(..., {\it COMM\_REDUCE})} so that each group has the
1-D distributed complete slab potential.
\item
Each process sends the gravitational potential on the slabs to other
processes in the group by calling {\tt MPI\_Alltoallv(..., {\it
  COMM\_SMALLA2A})}.  In the case of figure \ref{fig:relaymesh}, the
1-D distributed slab potential of eight processes are communicated to
nine processes.
\end{enumerate}

Using this method, we can reduce network congestion.  Here we present
the timing result for $4096^3$ FFT on 12288 nodes.  If we do not use
this method, the communication times for the conversion of the 3-D
distributed local density mesh to the 1-D distributed density slabs
and backward potential conversion were $\sim$10 and $\sim$3 seconds,
respectively.  With this method using three groups, these are reduced
to $\sim$3 and $\sim$0.3 seconds.  Thanks to our novel communication
algorithm, we achieve speed up more than a factor of four for the
communication.  
On the other hand, the calculation time of FFT itself was $\sim$4
seconds.  Thus, FFT became a bottleneck after the optimization 
of these communication parts.

One might think a 3-D parallel FFT library will improve the
performance further.  A 3-D parallel FFT library requires that
geometries of 3-D distributed density mesh in each process are the
same for all processes.  However, it is not practical to use similar
geometries for the domain decomposition of the particles and the mesh
in order to achieve good load balance. Thus, the communication with
the conversion of a 3-D distributed rectangular mesh to a 3-D
distributed regular mesh is needed and is likely to be more
complicated task than that of the combination of the 1-D parallel FFT
and relay mesh method since we have to consider additional two axes.
Although of course we consider to use such a library in the near
future, this novel technique should be also applicable for the
simplification of the conversion.

\section{Cosmological $N$-body Simulation}

\subsection{Calculation Setup}

We performed cosmological $N$-body simulations of $10240^3
(=1,073,741,824,000)$ dark matter particles on 24576 and 82944 nodes.
The latter is the full system of K computer.  This is the largest
cosmological $N$-body simulation has ever been done and is the first
gravitational trillion body simulation in the world.  The total amount
of memory required is  $\sim$200TB.  
The number of PM mesh was $N_{\rm PM}
= 4096^3$.  The cutoff radius for the short-range force $r_{\rm cut}$
are set to $r_{\rm cut} = 3/N_{\rm PM}^{1/3} \sim 7.32 \times 10^{-4}$,
where the side length of the simulation box is unity.

For the parallelization, the
number of divisions on each dimension is the same as that of physical
nodes of K computer.  These are $32 \times 54 \times 48$ ($32 \times
24 \times 32$) for the run on 82944 (24576) nodes. 
The number of divisions for the FFT processes is 
$16 \times 16 \times 16$ ($16 \times 8 \times 32$), 
the number of groups for relay mesh method is 18 (6).

We use a cube of the comoving size of 600 parsecs.  The corresponding
mass resolution is $7.5 \times 10^{-12}$ solar masses.  The smallest
dark matter structures are represented by more than $\sim 100,000$
particles.  The initial condition was constructed as positions and
velocities of particles represent the initial dark matter density
fluctuations with the power spectrum containing a sharp cutoff
generated by the free motion of dark mater particle (neutralino) with
a mass of 100GeV \cite{Green2004}.  The cosmological parameters
adopted are based on the concordance cosmological model
\cite{Komatsu2011}.

In this project, we focus on the dynamics of the smallest dark matter
structures at early Universe. We integrate the particle motion from
redshift 400 to $\sim$31.  For the time integration, we adopted the
multiple stepsize method \cite{Skeel1994}\cite{Duncan1998}.  The one
simulation step was composed by a cycle of the PM and two cycles of
the PP and the domain decomposition.   Figure
\ref{fig:snapshot} shows the snapshots of the simulation with 16.8G
particles.

\begin{figure*}[!t]
\centering
\includegraphics[width=16cm]{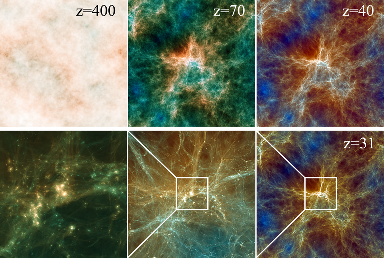}
\caption{
The distribution of dark matter of the 16.8G particles
  simulation at redshift 400 (initial), 70, 40, and 31.  The width of
  each image corresponds to 600 comoving parsecs.  Bottom-left and
  bottom-middle images are enlargements of the image of $z=31$. The
  sizes correspond to 37.5 (bottom-left) and 150 (bottom-middle)
  comoving parsecs. }
\label{fig:snapshot}
\end{figure*}

\subsection{Performance}

At the time of writing, the simulation is still running on 24576 nodes
of K Computer (corresponds to $\sim$30\% of the full system).  We had
an opportunity to measure the performance using the full system for
only a limited time.  The number of particle-particle interactions per
step averaged in the last five steps is $\sim$$5.3 \times 10^{15}$.
The calculation time per step is 173.8 and 60.2 seconds for 24576 and
82944 nodes, respectively.  Thus, the average performance on 24576 and
82944 nodes are 1.53 and 4.45 Pflops. The latter is $\sim$1.44 times higher
than that of the Gordon-Bell peak performance winner of last year
\cite{Hasegawa2011}.  Here, we use the operation count of 51 per
interaction following the description in subsection \ref{sec:kernel}.
The measured efficiency reaches 49 and 42\%.  It is important to keep
in mind that the performance is underestimated since we use only the
particle-particle interaction part to estimate the performance.  The
performance analysis with the Fujitsu sampling profiler shows the
efficiency a few percent higher since it counts all floating-point
operations.

Table \ref{tab:cost} gives the breakdown of the calculation cost per
step, and the performance statistics.  We can see that the short-range
part achieves near ideal load balance.  If we focus on the only force
calculation cycle, it achieves 71\% efficiency.  As seen in subsection
\ref{sec:kernel}, this value is equivalent to 95\% efficiency since
the theoretically maximum efficiency is 75\%.  On the other hand, the
long-range part shows load imbalance since FFT is parallelized for
only one axis.  The number of the FFT processes was $4096$, which is
smaller than that of all processes we used.  As a result, the
calculation cost of FFT is the same in the simulations on 24576 and
82944 nodes.  However, since the calculation cost of the long-range
part is minimized by our novel relay mesh method, the high performance
and excellent scalability is achieved even with the simulation on
82944 nodes.

Note that the average length of the interaction list
($\langle{N_j}\rangle \sim 2000$) is about 6 times smaller than that
of the previous Gordon-Bell winner who used a GPU cluster
\cite{Hamada2009}.  There are two reasons for this difference. 
The first reason is the difference of the boundary condition.  Hamada
et al. (2009) adopted the open boundary condition, and used the tree
algorithm for the entire simulation box.  On the other hand, we
adopted the periodic boundary condition and used the PM algorithm for
the long-range force.  
The $\log N$ term for our simulation is smaller that that of Hamada et al. (2009) 
because of the cutoff. The second reason is 
that they used large group size to achieve high performance 
on a PC with GPU.
The optimal value of $\langle N_i \rangle$ is
around $\sim$100 for K computer, and $\sim$500 for GPU cluster
\cite{Hamada2009}.

\begin{table}[!t]
\renewcommand{\arraystretch}{1.3}
\caption{ Calculation Cost of each part per step and the performance
  statistics.  One step is composed by a cycle of PM (long-range part)
  and two cycles of PP (short-range part) and domain decomposition.
  We used $N=10240^3$ particles.  }
\label{tab:cost}
\centering
\begin{tabular}{|lcc|}
\hline
$p$ (\#nodes) & 24576 & 82944 \\
$N/p$ & 43690666 & 12945382 \\ 
\hline\hline
PM(sec/step) & 9.28 & 6.74 \\
\quad density assignment &  1.44 &  0.44 \\
\quad communication  & 2.01 & 1.50 \\
\quad FFT &  4.06 & 4.17\\
\quad acceleration on mesh & 0.13 & 0.13 \\
\quad force interpolation &  1.64 & 0.50 \\ \hline
PP(sec/step) &  152.10 &  45.82 \\
\quad local tree &  4.00 & 1.26\\
\quad communication &  3.70 & 2.02 \\
\quad tree construction &  3.82 & 1.52 \\
\quad tree traversal &  17.17 &  4.60 \\
\quad force calculation &  122.18 &  35.72\\ \hline
Domain Decomposition(sec/step) &  6.28 &  5.38\\
\quad position update &  0.28 &  0.08\\
\quad sampling method &  2.94 &  3.80\\
\quad particle exchange &  3.06 &  1.50\\
\hline\hline
Total(sec/step) &  173.84 & 60.20\\ 
\hline\hline
$\langle{N_{i}}\rangle$ & 115 & 116 \\
$\langle{N_{j}}\rangle$ & 2346 & 2328\\
\#interactions/step & 5.35 Peta & 5.30 Peta\\
measured performance & 1.53 Pflops & 4.45 Pflops \\
efficiency & 48.7\% & 42.0\% \\
\hline
\end{tabular}
\end{table}

\section{Conclusion}

We present the performance results of the gravitational trillion-body
problem on the full system of K computer.  This is the largest
astrophysical $N$-body simulation and is the first gravitational
$N$-body simulation with one trillion particles.  This simulation is a
milestone that helps us to address the nature of the dark matter
particles, which is one of the long-standing problem in astrophysics
and particle physics.

The average performance achieved is 4.45 Pflops, which is 1.44 times
higher than that of the Gordon-Bell peak performance winner of last
year \cite{Hasegawa2011}.  The efficiency of the entire calculation
reaches 42\%.  The efficiency of the gravity kernel is 72\%.  These
high efficiency is achieved by a highly optimized gravity kernel for
short-range force calculation on the HPC-ACE architecture of K
computer and by developing a novel communication algorithm for the
calculation of long-range forces.  Our implementation enables us to
perform gravitational $N$-body simulations of one-trillion particles
within practical time.

We will further continue the optimization of our TreePM code.  The
current bottleneck is FFT.  We believe that the combination of our
novel relay mesh method and a 3-D parallel FFT library will
significantly improve the performance and the scalability.  We aim to
achieve peak performance higher than 5 Pflops on the full system of K
computer.

\section*{Acknowledgment} 
We thank RIKEN Next-Generation Supercomputer R\&D Center and Fujitsu
Japan at the RIKEN AICS (Advanced Institute for Computational Science)
for their generous support. Part of the results is obtained by early
access to the K computer at the RIKEN AICS.  K computer is under
construction, the performance results presented in this paper are
tentative.  The early development of our simulation code was
partially carried out on Cray XT4 at Center for Computational
Astrophysics, CfCA, of National Astronomical Observatory of
Japan. This work has been funded by MEXT HPCI STRATEGIC PROGRAM.

\bibliographystyle{IEEEtran}


\end{document}